\documentclass[]{spie}  

\usepackage{amsmath,amsfonts,amssymb}
\usepackage{graphicx}
\usepackage[colorlinks=true, allcolors=blue]{hyperref}
\usepackage{float}

\title{Visible camera cryostat design and performance for the SuMIRe Prime Focus Spectrograph (PFS)}

\author[a]{Stephen A. Smee}
\author[b]{James E. Gunn}
\author[a]{Mirek Golebiowski}
\author[a]{Stephen C. Hope}
\author[c]{Fabrice Madec}
\author[d]{Jean-Francois Gabriel}
\author[b]{Craig Loomis}
\author[c]{Arnaud Le Fur}
\author[c]{Kjetil Dohlen}
\author[c]{David Le Mignant}
\author[a]{Robert Barkhouser}
\author[b]{Michael Carr}
\author[a]{Murdock Hart}
\author[e]{Naoyuki Tamura}
\author[e]{Atsushi Shimono}
\author[f]{Naruhisa Takato}
\affil[a]{Johns Hopkins University, Department of Physics and Astronomy, 3701 San Martin Drive, Baltimore, MD 21218, USA}
\affil[b]{Princeton University, Department of Astrophysical Sciences, Princeton, NJ 08544, USA}
\affil[c]{Laboratorie d'Astrophysique de Marseille, Rue Frederic Joliot Curie, 13013 Marseille, France}
\affil[d]{Winlight Systems, 135 Rue Benjamin Franklin, 84120 Pertuis, France}
\affil[e]{Kavli Institute for the Physics and Mathematics of the Universe (WPI), The University of Tokyo, 5-1-5, Kashiwanoha, Kashiwa 277-8583, Japan}
\affil[f]{Subaru Telescope, National Astronomical Observatory of Japan, 650 North A'ohoku Pl. Hilo, HI 96720, USA}
\authorinfo{Further author information: (Send correspondence to Stephen A. Smee)\\ E-mail: smee@jhu.edu, Telephone: 1 410 516 7097}

\begin{document} 
\maketitle

\begin{abstract}
We describe the design and performance of the SuMIRe Prime Focus Spectrograph (PFS) visible camera cryostats.   SuMIRe PFS is a massively multi-plexed ground-based spectrograph consisting of four identical spectrograph modules, each receiving roughly 600 fibers from a 2394 fiber robotic positioner at the prime focus.  Each spectrograph module has three channels covering wavelength ranges 380~nm -- 640~nm, 640~nm -- 955~nm, and 955~nm -- 1.26~um, with the dispersed light being imaged in each channel by a f/1.07 vacuum Schmidt camera.  The cameras are very large, having a clear aperture of 300~mm at the entrance window, and a mass of $\sim$280~kg.  In this paper we describe the design of the visible camera cryostats and discuss various aspects of cryostat performance. 
\end{abstract}


\section{INTRODUCTION}
\label{sec:intro}  
The SuMIRe Prime Focus Spectrograph~\cite{Tamura} (PFS) is a large spectroscopic survey instrument consisting of four identical spectrograph modules, each receiving roughly 600 fibers from a robotic fiber positioner at the telescope prime focus.  Each spectrograph module has three channels, two visible and one near infrared, covering the wavelength ranges 380~nm -- 640~nm, 640~nm -- 955~nm, and 955~nm -- 1.26~$\mu$m. The cameras for each channel are almost identical vacuum Schmidts.  Spectra are imaged onto detectors with a 4k~x~4k format and 15~$\mu$m pixels.  A pair of 2k x 4k Hamamatsu CCDs comprise the detector array for the two visible channel cameras~\cite{Pascal2014}. For the near infrared channel camera~\cite{2014SPIE.9147E..2VS} we use the Teledyne H4RG-15, 4k~x~4k 15~$\mu$m pixels.

The PFS project is a collaboration, with hardware contributions coming from a number of institutions around the world. In the case of the visible cameras, the optics and their mounts are being developed, fabricated and assembled in France, by the Laboratoire d'Astrophysique de Marseille (LAM) and Winlight Systems (WS).  The cryostat is being developed, fabricated, and assembled by Princeton University (PU) and Johns Hopkins University (JHU) in the United States.  The split responsibility (optics and cryostat) has certainly complicated the design interfaces, which we discuss in Sec.~\ref{sec:mech_design}. 

In this paper we describe the design and performance of the visible camera cryostat, which houses the optics and cools the detector.  We discuss the mechanical design, the thermal design, and the control system, as well as results from performance tests conducted at Johns Hopkins and the Laboratorie d'Astrophysique de Marseille.

\section{CRYOSTAT MECHANICAL DESIGN} 
\label{sec:mech_design}
The PFS spectrograph visible channels use large, fast (f/1.07), vacuum Schmidt cameras having four optical elements: a dual-element corrector, a Schmidt primary, and a field-lens in front of the detector. The optics and detector are housed in a vacuum, defined by the body of the camera, which we refer to as the \emph{cryostat}.  The corrector and Schmidt primary, a Mangin mirror, along with their supports, operate at ambient temperature. The detector is cooled to a nominal operating temperature of 163~K by a Stirling-cycle cryocooler.  Figure~\ref{fig:Red_Camera} shows the arrangement for the red channel camera; the blue camera is nearly identical.  Each camera, including the mount, has a mass of roughly 280~kg.  The entrance aperture has a diameter of 320~mm, and the diameter of the cryostat body is 667~mm.  These are very large cameras.

The cryostat body consists of four main components stacked end-to-end: the front bell, the front ring, the rear tube, and the rear cover.  See Fig.~\ref{fig:red_cryostat_exploded}. All of these components are fabricated from 6061-T6 aluminum.  Adjacent sections are bolted together and sealed by a Viton O-ring, and the O-ring groove has a dovetail cross-section, which prevents the O-ring from falling out during assembly/disassembly.  A boss on the back side of the front bell centers the front ring, and orientation is set by a single Delrin pin.  The rear cover is registered to the rear tube in a similar fashion. However, the front ring is registered to the rear tube differently; two Delrin alignment pins are used for alignment, providing centration and clocking. This minimizes the wall thickness.  The outer surface of the cryostat body is color-coded for the respective channel; red anodize for the red channel, and blue anodize for the blue channel.  The only physical difference between the red and blue cryostats, besides the color, is the length of the thermal bars (see Sec.~\ref{sec:thermal_design}), which are slightly longer in the blue camera

All of the cryostat vacuum, thermal, and control hardware is mounted on the rear cover.  Each of these systems is discussed in detail in the sections that follow.  The thermal spreader, thermal bars, and thermal straps, all mounted on the inside of the rear cover, carry heat from the detector to the cryocooler, which is mounted on the outside of the cover.  Two ion pumps, a turbo pump, and vacuum gauge are also mounted on the outside of the rear cover.  The cryostat control electronics reside inside a sealed and actively cooled enclosure called the \emph{pie-pan}, which is mounted on the outside of the rear cover as well, just behind the vacuum hardware.

\subsection{Electrical Feedthroughs}
All electrical signals into and out-of the cryostat are routed through one of two feedthrough modules attached to the front ring. One module, dubbed simply the \emph{feedthrough} module, is dedicated to temperature sensors, focus motors, limit switches, and heaters. The second module, dubbed the \emph{FEE} module, is the Front-End Electronics (FEE) for the CCD readout.  The FEE sets bias voltages, monitors voltages, and does the analog-to-digital conversion of the video signals coming from the CCDs.  These two modules are cabled to the pie-pan on the outside, and to their various respective components on the inside. 

\subsection{Interfaces}
\begin{figure}
    \centering
    \includegraphics[width=16cm]{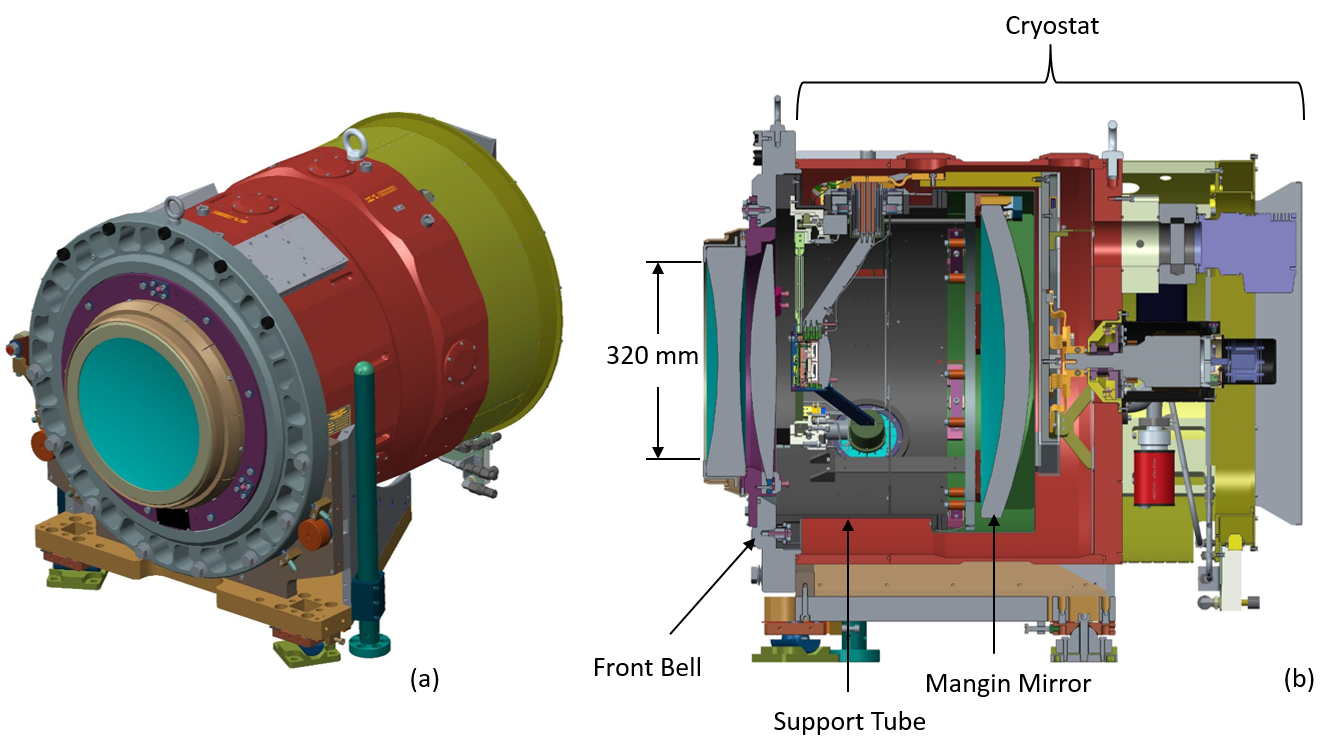}  
    \caption{(a) Rendering of the PFS red channel camera.  (b) Cross-section of the red camera highlighting the internal optics and optics support.}
    \label{fig:Red_Camera}
\end{figure}

As mentioned in the introduction, the responsibility for the cryostat belongs to PU and JHU.  And the responsibility for the camera optics belongs to LAM and WS.  It is clear from Fig.~\ref{fig:Red_Camera} that the front bell is both integral to the optics as well as the cryostat.  This complicates the design interface, but in practice has proved not to be an issue.  LAM and WS construct the front bell and integrate their optics to it, while PU and JHU construct the remainder of the cryostat, then ship it France where it is integrated to the optics assembly.  For testing at JHU, a \emph{dummy} front bell is used in place of the real front bell supplied by WS.

Internal to the cryostat, the interfaces are more tricky.  For example, thermal sensors mounted within the support tube belonging to LAM and WS, are the responsibility of LAM and WS.  Sensors attached to the thermal spreader and cooler are the responsibility of PU and JHU.  Reading the sensors, all of them, is the responsibility of PU and JHU.  The focus mechanism, and its harness, are also the responsibility of PU and JHU, even though these components are internal to the support tube.

\begin{figure}
    \centering
    \includegraphics[width=17cm]{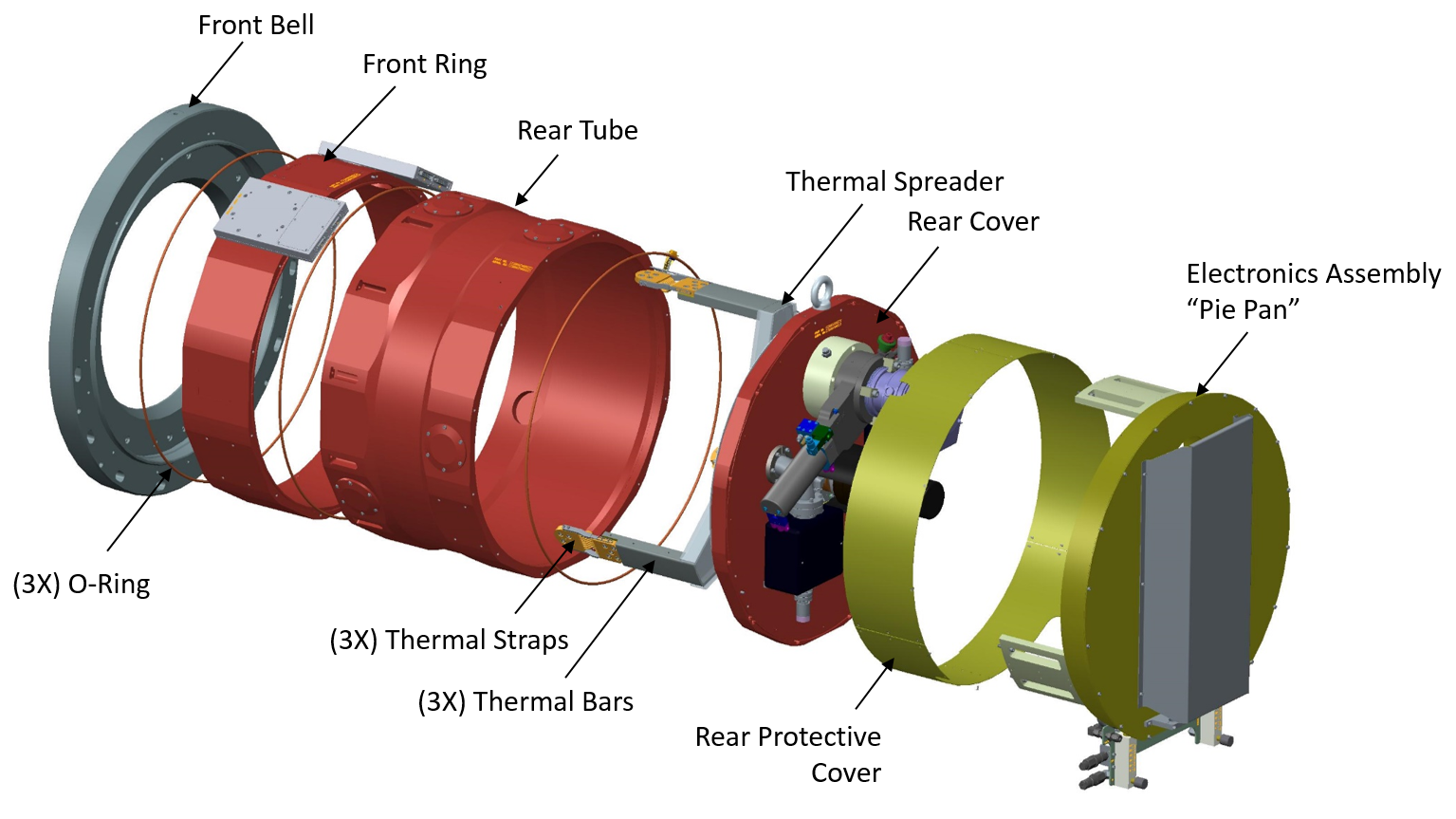}  
    \caption{Exploded view of the red cryostat.  Components stack end-to-end. From front (left) to back (right) the main components are the front bell, the front ring, the rear tube, and rear cover.  The control electronics, or \emph{pie-pan}, attach to the back of the rear cover.}
    \label{fig:red_cryostat_exploded}
\end{figure}

The thermal interface between our institutions resides at the connection of the thermal straps highlighted in Fig.~\ref{fig:red_cryostat_exploded}.  Specifically, where the straps connect to the thermal rods inside the support tube; see Sec.~\ref{sec:thermal_design}.  PU and JHU are responsible for the thermal strap and all components along the thermal path back to the cryocooler.  LAM and WS are responsible for the thermal rod and all components along the thermal path leading to the detector.

\section{THE VACUUM SYSTEM}
\label{sec:vacuum}
The cryostat has an internal diameter of 625~mm and is 488~mm long for a volume of approximately 150 liters. The estimated surface area internal to the cryostat is 19~m$^2$, including the optics and optics supports. All ports and interfaces are sealed with vacuum-baked Viton O-rings, totaling roughly 13~m of O-ring length.  Under steady state operation, permeation through the O-rings will dominate, leading to a gas load of approximately 13 x $10^-6$ Torr$\cdot$liter/sec.  

\subsection{Steady State Operation}
For steady state operation, the cryostat is pumped using two Agilent VacIon plus 20 ion pumps, roughly 3$x$ the expected gas load at 1 microTorr.  The pumps are mounted on the rear cover of the cryostat, attached to 90$arcdeg$ bent-tube elbows for compactness, and to mitigate the glow of the ion pump as seen from inside the cryostat; see Fig.~\ref{fig:red_camera_rear}. Custom KF support rings, consisting of a coarse wire mesh welded to a stock ring, are installed where the bent tube attaches to the rear cover.  The mesh serves to discharge any plasma created by a high-pressure start-up of the ion pumps; an unlikely event but not out of the realm of possibility, and were it to happen without protection could destroy the CCDs.  Pressure is monitored by a Pfeiffer MPT 200 Pirani/coldcathode combination gauge.

\subsection{Evacuating the Cryostat}
\begin{figure}
    \centering
    \includegraphics[width=12cm]{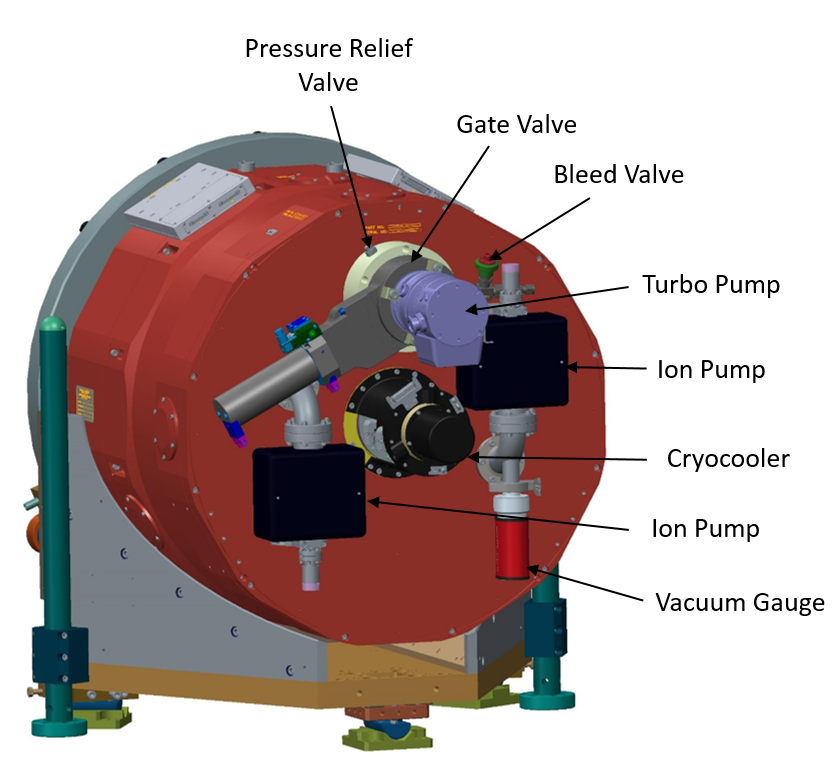}  
    \caption{Rendered image showing the vacuum hardware mounted to the rear cover of the red camera cryostat. Hardware includes two ion pumps, a turbo pump separated from the cryostat by a gate valve, a vacuum gauge, a bleed valve, and a pressure relief valve to avoid over-pressure.  The cryocooler is mounted in the center of the rear cover.}
    \label{fig:red_camera_rear}
\end{figure}

An Edwards EXT75DX 60 liter/sec turbo pump is used to evacuate the cryostat.  The pump is mounted atop an ISO63 VAT gate valve on the rear cover of the cryostat, just above the cryocooler; again see Fig.~\ref{fig:red_camera_rear}. A riser between the gate valve and the rear cover allows the valve to clear the ion pump bent-elbow, and also provides a convenient location for a bleed valve -- used to bring the cryostat back up to ambient pressure -- and a pressure relief valve -- used to avoid an over-pressure condition on the corrector window during a nitrogen purge.  The corrector window is not susceptible to failure, but an over-pressure condition could separate the corrector window from the O-ring seal; a situation we would like to avoid. 

\section{CRYOSTAT THERMAL DESIGN}
\label{sec:thermal_design}
\begin{figure}
    \centering
    \includegraphics[width=16cm]{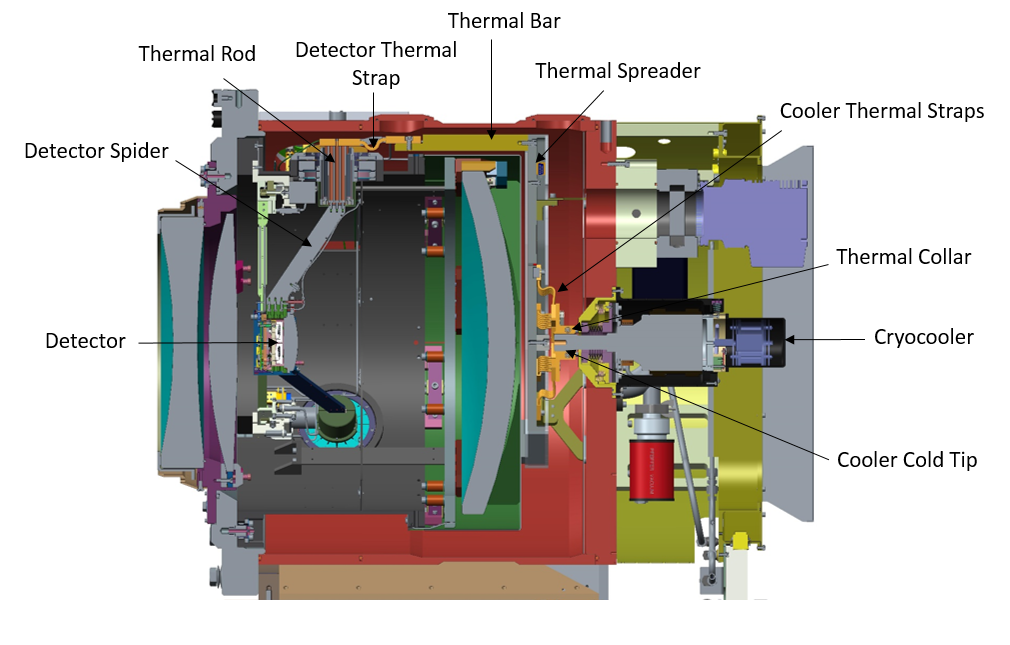}  
    \caption{Rendered cross-section of the red cryostat highlighting the thermal path from the detector to the cryocooler.  Starting at the detector, heat flows to the detector mounting box, then flows through three parallel paths, each path consisting of a detector spider, a thermal rod, a detector thermal strap, and a thermal bar.  These three paths terminate a the thermal spreader, which conducts all of the heat through eight thermal straps to a thermal collar clamped to the cold tip of the cryocooler.}
    \label{fig:red_cryostat_thermal}
\end{figure}

The sole purpose for the visible cryostat is to cool the detectors to their operating temperature; nominally 163~K. The evacuated volume eliminates convective heat transfer limiting losses to radiation and conduction.  In the PFS visible cameras, heat is removed from the detector using a single Sunpower Cryotel GT mounted to the rear of the cryostat.  Figure~\ref{fig:red_cryostat_thermal} shows the placement of the cooler, as well as the components that make up the thermal path.  Starting at the detector, heat flows to the detector mounting box, then flows through three parallel paths, each path consisting of a detector spider, a thermal rod, a detector thermal strap, and a thermal bar.  These three paths terminate a the thermal spreader, which conducts all of the heat through eight wire-rope thermal straps to a thermal collar clamped to the cold tip of the cryocooler. The detector mounting box, detector spider, thermal rod, and thermal straps are all made of oxygen free copper.  The thermal bars and thermal spreader are made of 1100 series aluminum. The thermal collar is made of AlBeMet 162, which has a very low density and high thermal conductivity.  AlBeMet was used here specifically because of its low density, the concern being that the lateral load specification on the cooler tip could be exceeded in the unlikely event of a mechanical shock to the camera; possible during shipping for example.

\subsection{Radiative Thermal Losses: Cryostat only}
With the exception of surfaces associated with the thermal path from the detector to the cooler, all surfaces within the cryostat operate at ambient temperature, 5~C in the case of PFS.  To minimize radiative coupling from cryostat walls to the cold components, the interior surfaces of the front ring and rear tube are polished, and a polished aluminum shield is attached to the inside of the heavily light-weighted rear cover. To further reduce radiative losses, a fifteen-layer Multi-layer insulation (MLI) blanket surrounds the thermal spreader.  And the thermal bars are polished. 

Assuming a 3\% emissivity for the MLI and 4\% for the polished surfaces, the total radiative loss from the cryostat, exclusive of the optics, is approximately 7.1~W. 

\subsection{Conductive Thermal Losses: Cryostat Only}
The only conductive losses from the cryostat are through the G10 struts that support the thermal spreader off the rear cover, and the wiring that routes through the two feedthroughs. We estimate the loss through the G10 struts to be approximately 0.4~W.  We have not estimated the losses through the wires.  Based on estimates from previous instruments the wires will contribute very little to the heat load.

\subsection{Thermal Losses from the Optics}
While the optics and their supports are not considered part of the cryostat, the thermal losses are obviously very important to the cryostat design.  Based on detailed analysis by Winight Systems, the estimated heat load from the optics is approximately 7~W. 

Therefore, the total estimated heat load is $\sim$14.5~W.

\subsection{The Cryocooler}
Each visible camera cryostat is cooled by a single Sunpower Cryotel GT Stirling cycle cooler.  The Sunpower cooler was chosen because it has several key advantages over other technologies including very long lifetime, a compact form factor, and greatly reduced heat dissipation to the ambient as compared to traditional compressor-based systems. At full power, the GT has a lift of 15~W at 77~K, 22~W at 100~K, and 27~W at 120~K.  For our expected heat load and thermal impedance, we anticipate a cooler tip temperature between 100-110~K, with the cooler running at roughly 60\% power.

The only concern with the Sunpower cooler is vibration; a significant consideration for astronomical instruments. The self-contained unit vibrates due to the translating piston internal to the unit.  With the factory supplied passive damper, the unit vibrates at at the level of 400~mG at 60 Hz, and to a lesser degree at higher-order harmonics of 60 Hz. This is a bit much for our application.  However, Sunpower has recently developed an active damping control system for the Cryotel units.  And when combined with a properly designed mount, (see Sec.~\ref{sec:sunpower_mount}) we have found that the vibrational energy transmitted to the cryostat is negligible.  Recent test results show vibration levels of order 1 mG at the cryostat with the cooler running at full power (the measurements were made with the accelerometer attached to the end of one of the three thermal bars).   

Heat generated by the coolers is removed by facility glycol coolant.  Dissipation from the Cryotel GT when operating at full power is roughly 250~W, and with a glycol flow rate of 1~l/min the expected temperature rise of the body of the cooler is of order 10~K, small enough to produce only a minimal amount of convective heat leak to the dome.  From experience, we know that only a small portion of the body of the cooler reaches this elevated temperature; most of the surface runs at ambient. 

\subsection{The Sunpower Cryotel GT Mount}
\label{sec:sunpower_mount}
\begin{figure}
    \centering
    \includegraphics[width=16.0cm]{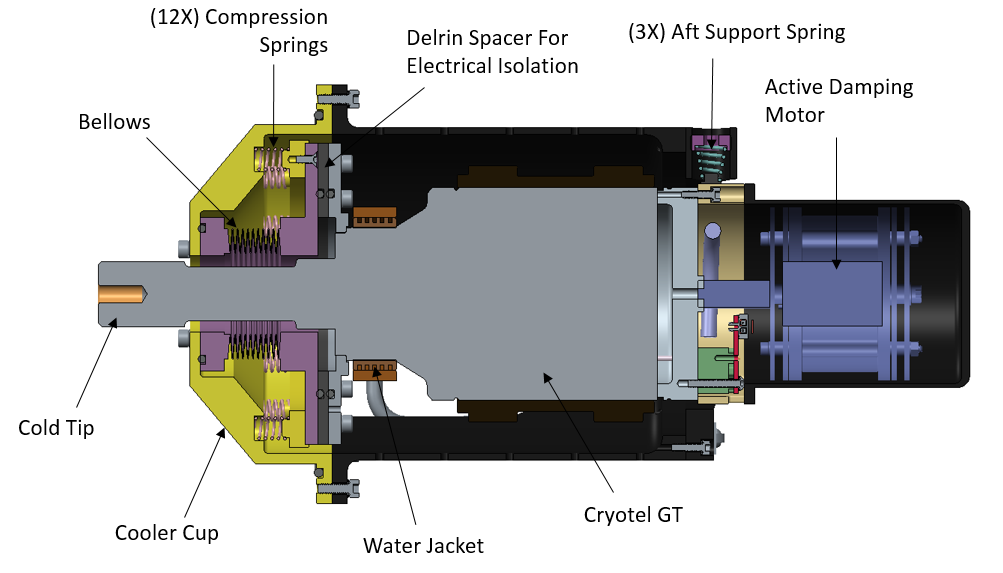}  
    \caption{Rendered cross-section of the Sunpower Cryotel GT mount developed for PFS.  The front half of the cooler is supported on-axis by twelve compression springs. A compliant bellows seals the cooler front flange to the base of the cooler cup, and also provides lateral support.  Support at the rear of the cooler is provided by three conical springs.  The active cancellation motor is attached at the rear of the unit.}
    \label{fig:CryotelGT}
\end{figure}

Forces transferred to the cryostat are minimized if the cooler is mounted in a compliant fashion. To achieve this, a custom mount was designed that allows the cooler to float on springs.  Details of the design are shown in Fig.~\ref{fig:CryotelGT}.  The front half of the cooler is supported on-axis by twelve compression springs. A compliant bellows seals the cooler front flange to the base of the cooler cup, and also provides lateral support.  Support at the rear of the cooler is provided by three conical springs.  The active cancellation motor is attached at the rear of the unit. We are grateful to Gerry Luppino for sharing the bellows mounting concept used here, which served as the starting point for this design.

\subsection{Thermal Straps}
The thermal connection between the cooler and the thermal spreader is made using eight copper wire-rope thermal straps, provided by Thermal Space Limited (\url{http://thermal-space.com/}).  Compliant copper straps of similar construction are also used to connect the thermal bars to the thermal rod. In each case, the main motivation for using these straps is vibration isolation.  The straps themselves consist of two solid copper lugs separated by a set of fine gauge wire copper rope.  They are very flexible. The rope/lug termination is a swaged connection, so the flexibility extends all the way to the lug. Each of the eight cooler straps uses 3~mm diameter rope and has a conductance of roughly 0.18~W/K.  Each of the three detector straps uses 5~mm diameter rope and has a conductance of approximately 0.78~W/K.

\begin{figure}
    \centering
    \includegraphics[width=10.0cm]{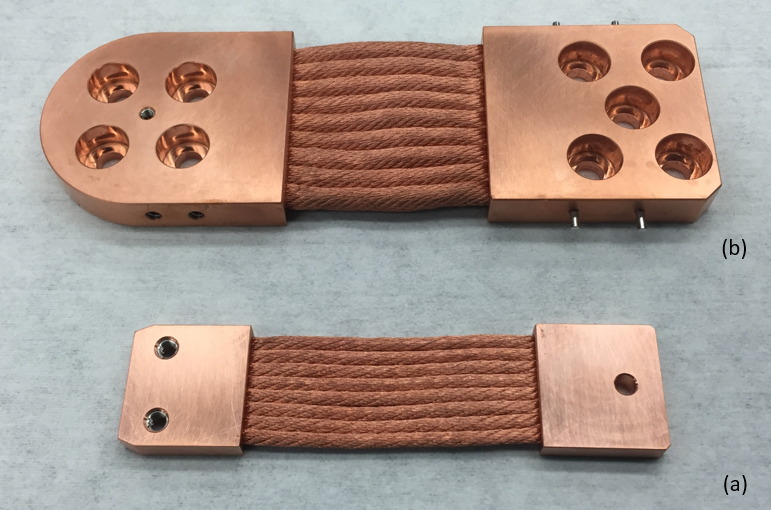}  
    \caption{Photograph of the two types of thermal straps used in the PFS visible camera cryostat: (a) the cooler strap, which connects the cooler collar to the thermal spreader, and (b), the detector thermal strap, which connects the thermal bar to the thermal rod.}
    \label{fig:straps}
\end{figure}

\subsection{Heaters and Heater Control}
Within the camera there exists two heater circuits.  One for control of the detector temperature.  And a second for warming up the cyrostat. 

The detector heater circuit consists of four 220~Ohm heaters bonded into the thermal straps that conduct heat from the detector mounting plate to the detector box; see Ref.~\citenum{Gunn} for a more detailed description of the detector mount.  The heaters are configured in pairs.  Each pair is wired in series, and the pairs in parallel.  Hence, the net resistance is 220~Ohm.  The heaters are controlled by the temperature control module (see Sec.\ref{sec:tempboard}) from one of two PID channels.  A maximum of 2.3~W can be applied to the heaters in this configuration.  To prevent an over-temperature condition, a Klixon 3BT-L6 thermostat is attached to the CCD mount plate.  The normally closed thermostat has a nominal open temperature of 21~C.

For cryostat warmup, a set of three Kapton film heaters are bonded to the thermal spreader.  The three 208~Ohm heaters are wired in parallel for a net resistance of 70~Ohm.  This circuit is also under PID control from the temperature control module. A maximum of 7~W is permissible with this configuration.  To prevent an over-termperature condition, a Klixon 5BT thermostat is attached to the spreader and wired in series with the power lead.  The 5BT-5 also has an open temperature of 21~C.

\subsection{Temperature Sensors}
We have opted to use RTDs (resistance temperature detector) rather than diodes for monitoring temperatures within the cryostat; mainly due to cost. Three flavors are used.  All but two of the RTDs are 1000~Ohm RTDs (Honeywell HEL-705-U-1-12).  The cryocooler tip temperature is measured by a 100~Ohm RTD from Lake shore Cryotronics (model number XPT-111-45).  This sensor is mounted to the thermal collar and supplied by Sunpower.  For the cooler reject temperature (the hot side of the cooler) we use a Honeywell 100~Ohm RTD (model HEL-705-T-1-12) reject temperature. 

The cryocooler RTDs, both tip and reject temperature, are read out by the cryocooler controller.  All other sensors are read by the temperature control module, which is discussed in some detail below.

\section{CRYOSTAT CONTROL}
\label{sec:control}
\begin{figure}
    \centering
    \includegraphics[width=16.0cm]{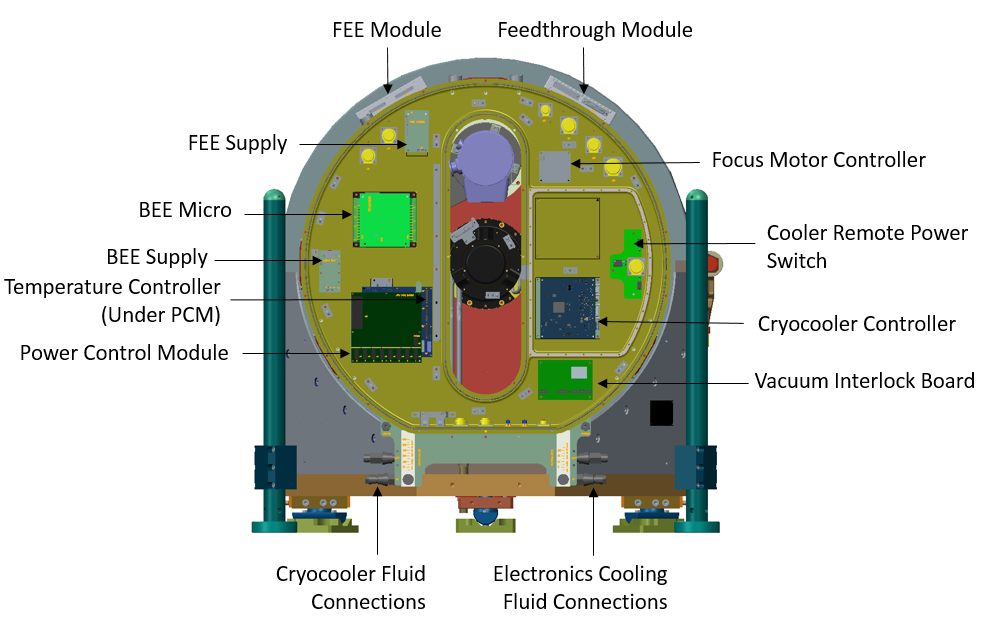}  
    \caption{Rendered image of the rear of the cryostat with the electronics enclosure cover removed to reveal the components inside.  This single enclosure, the pie-pan, includes all of the control electronics for controlling the pumps, cryocooler, focus motors, as well as readout of the temperature sensors, pressure gauge, and detector readout.}
    \label{fig:piepan}
\end{figure}

Control of the cryostat, including the pumps, cryocooler, focus motors, as well as readout of temperature sensors, the pressure gauge, and detector readout are all handled by the electronics inside the electronics enclosure at the rear of the cryostat; the pie-pan.  This single location makes for a convenient, and accessible, way to package the electronics. Figure~\ref{fig:piepan} shows a rendered image of the enclosure.  The contents, and their function, are described in some detail below.  Power and communications from the instrument control rack enter the enclosure at a bulkhead on the bottom of the pie-pan, with the exception of the 48~V supply for the cryocooler controller; for noise reasons, that supply enters the enclosure through a passthrough in the rear of the pie-pan adjacent to the cooler controller.  Signals to the Front End Electronics (FEE) and the Feedthrough module are routed through two sets of connectors: one clustered around the top-left side of the pie-pan in the vicinity of the FEE supply, and the other clustered around the top-right side near the motor controller.  

The total power dissipated by the pie-pan electronics is approximately 48~W.  The heat is removed by two Lytron cold-plate heat exchangers on the outside of the enclosure: in Fig.~\ref{fig:piepan} one runs vertically from the BEE micro down to just beneath the Power Control Module; the other runs vertically as well and resides just under the cryocooler controller.  Fluid connections for these cold-plates, as well as for the cryocooler, are mounted on a bulkhead plate at the bottom of the enclosure.  Quick disconnect fittings facilitate quick, clean, connection of the coolant lines.  The coolant, a  glycol(40\%)/water mixture is temperature controlled and supplied by the Subaru facility. A flow rate of 0.25~l/min is used for the cryocooler, and a flow rate of 1~l/min is used for the electronics.

To avoid transmitting noise to the environment, the enclosure is sealed using spiral Flexi-Shield O-ring gaskets where the main cover seals against the enclosure.  The gaskets are a tin(90\%)/lead plated stainless steel spiral wrap over a 40 durometer silicone tubing.  The effective shielding is 100-120~dB at 1~GHz.

The cryocooler controller is also shielded within the enclosure; the controller resides in an enclosure \emph{within} the enclosure.  This is done to mitigate the risk of transmitting radio frequency interference from the switching supplies in the controller to the BEE micro, or the cabling that leads to the Front End Electronics.  The lid to this enclosure (visible in Fig.~\ref{fig:piepan_photo}) is sealed using an Ohmerics S1081 nickel/graphite silicone elastomer gasket.

\subsection{FEE Supply}
The FEE supply is a custom power supply board that provides DC voltages to the Front End Electronics.  The supply incorporates 7 switch mode, DC to DC, converters to produce the following voltages: 3.3~V, +5~V, -5~V, +12~V, -12~V, 24~V, and 54~V.  The FEE buffers each supply with high power supply rejection ratio (PSSR), low dropout (LDO) regulators, to reduce noise. Each supply is enabled and disabled by the power control module.

\subsection{BEE Micro}
The Back-End Electronics (BEE) is the detector readout and control microprocessor for the visible cameras.  It is a commercial device from RTD Embedded Technologies; specifically, their CMX34GSS615HR-2048 PC104e with PCI express bus processor board.  Atop the microprocessor board is a FPGA35S6045HR (FPGA35S6) PC104e FPGA board, which includes a Xilinx Spartan 6 FPGA, with custom programming capability.

The BEE performs a number of functions related to readout of the detector.  Those functions are: to provide all the necessary timing and control signals to clock analog data from two 2k~x~4k Hamamatsu CCDs; to provide all the necessary timing and control signals to synchronously read digital data from eight Analog Devices ADC7686 16 bit analog-to-digital converters located on the Front End Electronics board (FEE); to provide differential UART interface to a microcontroller located on the FEE; to provide memory accessible by both the FPGA and the Linux operating system to facilitate the storage and subsequent upload of captured images; and to provide Gigabit Ethernet to allow remote command, control, image transfer and telemetry.  All interfaces between the BEE and the FEE utilize Low Voltage Differential Signals (LVDS) with the exception of the RS232 Serial Port UART.

The BEE has been programmed to provide a number of clocking schemes.

For a full discussion of the PFS CCD readout electronics, including the FEE and pre-amp board, see Ref.~\citenum{2014SPIE.9154E..2GH}.

\subsection{BEE Supply}
The BEE supply is a custom built supply that incorporates 2 switch mode, DC to DC, converters to produce two voltages: 5.1~V and 5.25~V. The BEE only uses the 5.1~V supply. The 5.25~V supply is used for the near infrared camera, and rather than building two separate supplies we built one that provides both voltages.

\subsection{Power Control Module}
Power distribution and communication within the enclosure is handled by the Power Control Module (PCM).  It is a custom printed circuit board developed by JHU for the James Webb Space Telescope (JWST) photogrammetry system.  The board has a four port Ethernet switch for communication to other modules such as the temperature control module, and eight channels of RS485 communication and power switching and monitoring for peripherals operated by DC power on one of two busses (27.6~V battery backed, or 24~V volatile), such as the stepper motor controller. Hence, the board is essentially a programmable switch for many devices, and serves as a communication hub for those devices as well.  The PCM dissipates very little heat; approximately 1~W.

\subsection{Temperature Control Module}
\label{sec:tempboard}
Temperature monitoring and heater control is handled by the Temperature Control Module (TCM); also a custom built board with heritage from the JWST photogrammetry system. The TCM has 12 sensor channels and controls two PID heater channels at up to 100~W per channel.  A limiting resistor sets the heater channel power.  Therefore for low power applications such as detector thermal control, the power is easily reduced to a lower level.  The board communicates via the Ethernet protocol to the PCM, which resides just above it in the pie-pan. It has the same form-factor as the PCM board so the two boards can be stacked in a compact configuration, however in this implementation, with little gap between PCM and TCM, the position of the TCM is staggered to provide access to the heater and sensor connections.  The board dissipates approximately 5~W. 

\subsection{Focus Motor Controller}
The focus mechanism consists of three stepper-motor-driven levers that translate the detector up to $\pm150\mu$m with fine resolution.  Each motor is controlled from a single board, an off-the-shelf, four-axis, stepper motor controller from All Motion; model EZ4AXIS. The fully intelligent stepper motor Controller + Drivers runs four fully independent stepper motors at up to 1~A.

\subsection{Cooler Remote Power Switch}
The cryocooler controller operates from 48~VDC. A custom DPST (Double Pole Single Throw), Solid State Relay (SSR), controlled with a 24~V control signal from the PCM, allows remote switching of the cryocooler supply.  The design uses two N-Channel MOSFETs per channel to control each output. When energized, current initially flows through an inrush limiting thermistor with a nominal resistance of 10 Ohms, limiting current inrush to less than 5~A. After a short one second delay the thermistor is removed from the circuit by a second MOSFET resulting in negligible I$^2$R losses. All switching occurs on the high-side. Optocouplers provide isolation between the control input and the 48~V switch.

\subsection{Cooler Controller}
\begin{figure}
    \centering
    \includegraphics[width=10.0cm]{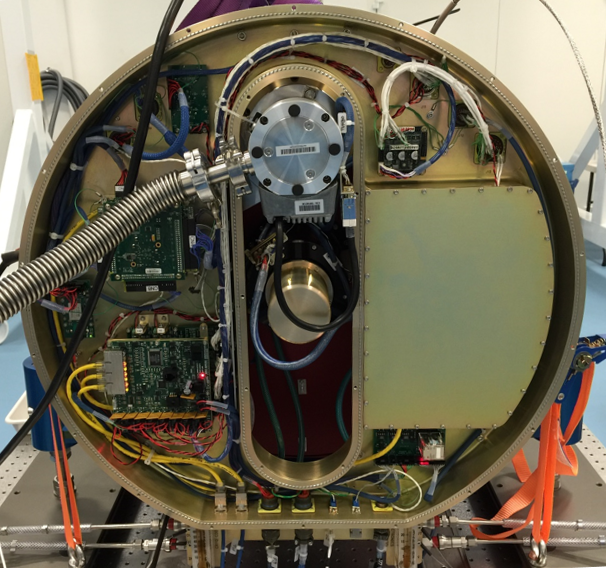}  
    \caption{Photograph of the red cryostat electronics enclosure with the main cover removed.  The cryocooler controller is encased beneath the panel on the right-hand side of the photo.}
    \label{fig:piepan_photo}
\end{figure}

The cryocooler controller was custom built for PFS by Sunpower. The design incorporates the electronics for their active cancellation system, and is based in large part on their standard commercial offering.  However, their standard controller was too large for our enclosure, so we had them re-design it to fit.  The controller is housed in a RF (radio frequency) sealed compartment on the right side of the pie-pan enclosure; see Fig.~\ref{fig:piepan}, and for a photograph Fig.~\ref{fig:piepan_photo}. The RF sealed volume is actually large enough for two controllers; we use a second controller in the NIR camera and wanted a common design.  The base of the controller is heat-sunk to the base of the pie-pan.  A Lytron cold-plate heat exchanger mounted on the outside of the enclosure, and directly beneath the controller, removes the waste heat; approximately 20~W.

\subsection{Vacuum Interlock Board}
The vacuum interlock board interfaces the PCM and the BEE to the turbo pump and gate valve. A 24~V control signal from the PCM enables the gate valve interlock. Once enabled, the gate valve can be opened with a control signal from the BEE. Once the turbo is up to speed, gate valve control is locked out. In the event the turbo fails, the gate valve will close automatically regardless of the state of the control signal. External control can only be regained by cycling the enable input. During normal operation, the gate valve will typically remain closed. A connector on the pie-pan input bulkhead provides a means of disabling BEE control, ensuring that the gate valve cannot be opened accidentally. The board also returns Gate Valve and Turbo status data to the BEE.

\section{PERFORMANCE}
\label{sec:performance}
The first visible red cryostat was shipped from Johns Hopkins University to the Laboratorie d'Astrophysique de Marseille in September 2015. Prior to shipment, the electrical, thermal, and vacuum performance was validated through a number of tests.  After delivery, the vacuum and thermal tests were repeated, and additional tests were conducted with the optics installed.  In this section we describe these tests in some detail and discuss what went right, and what went wrong.

\subsection{Vacuum and Thermal Performance Tests at JHU}
\label{sec:performance_JHU}
Vacuum leak-checking of the first red cyrostat began in late July, 2015. No leaks were found and we proceeded with initial operation of the vacuum system, albeit with a crude set of controls; at this point the pie-pan had not been completed.  The first pumpdown proceeded normally with the pressure dropping to 5~x~$10^-5$~Torr in 24 hours with the ion pumps off.  After 48 hours, the pressure had reached 2~x~$10^-5$~Torr, shortly after which the turbopump stopped due to an over-speed condition.  The pump was replaced and the pressure recovered quickly.  A little more than four days after the pumpdown started, the pressure had reached 1~x~$10^-5$~Torr, the ion pumps were started and the pressure dropped to roughly 5~x~$10^-6$~Torr. Pumping was halted and the cryostat was pressurized in preparation for thermal testing. We should note that the cause of the turbopump over-speed condition is not known and we have not had an issue with subsequent pumps.

By mid-August 2015, the pie-pan was installed and a thermal validation test was conducted with the cryostat under full electrical and software control. The goals of the thermal test were two-fold: one, to verify the thermal impedance between the cooler tip and the thermal bars; and two, to verify the heat load estimates discussed above.  To conduct this test, the temperature at the ends of the thermal bars was measured with a total of 7~W heat applied to the ends of the bars, with the cryocooler set to 100K. Factoring in the thermal impedance of the detector support system (i.e. thermal strap, thermal rod, and detector spider), the temperature of the thermal bars should be less than 140~K.  Based on the total estimated heat load, 14.5~W, we should be able to easily achieve this temperature with a fraction of the cooler power.

\begin{figure}
    \centering
    \includegraphics[width=17.0cm]{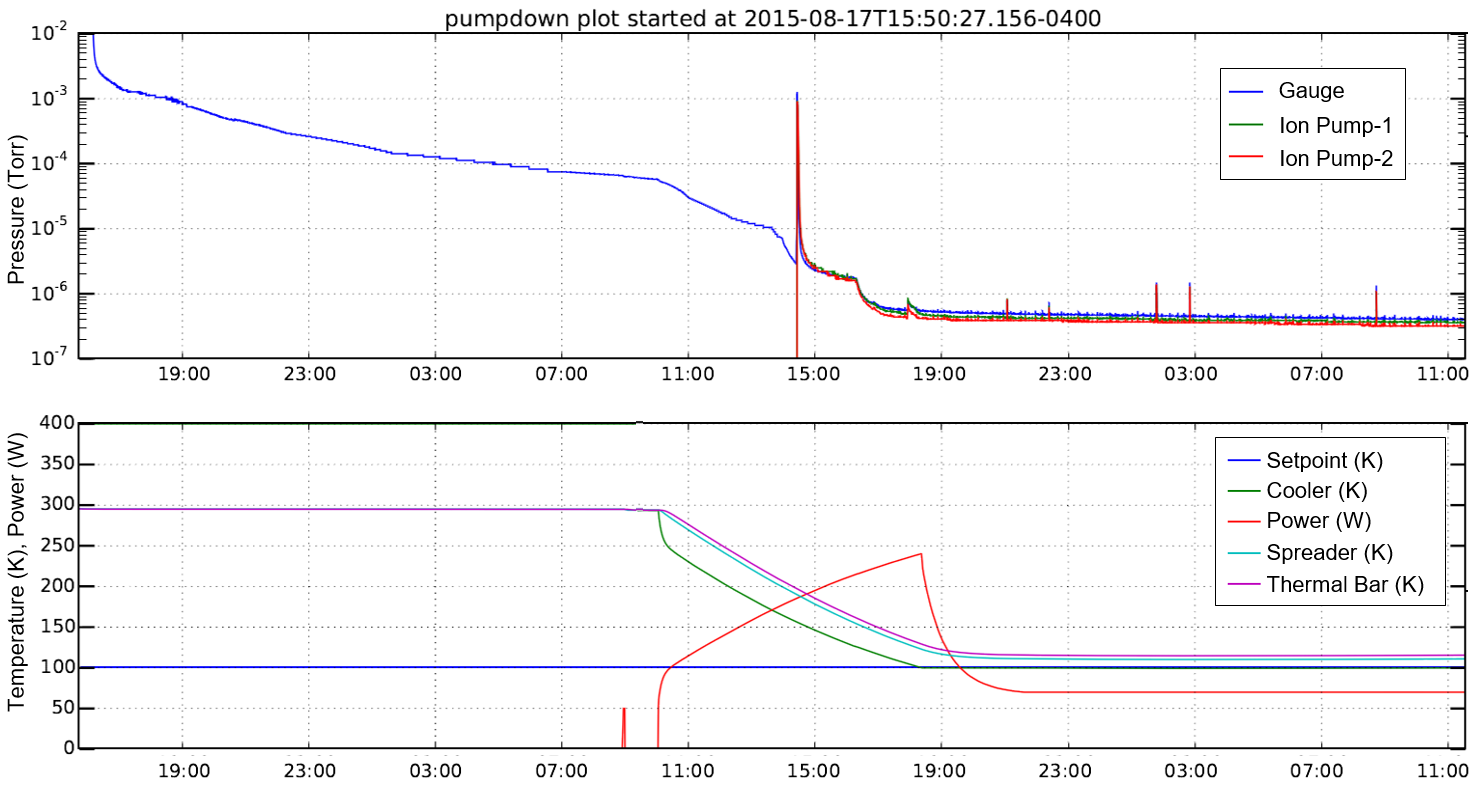}  
    \caption{Plot showing the cyrostat pressure (top panel), and temperatures and cooler power (bottom panel), for the cooldown of the first red cryostat. The optics were not installed for this test, and no heater power was applied to the thermal bars to simulate the ambient load produced by the optics.}
    \label{fig:cooldown_JHU}
\end{figure}

Figure~\ref{fig:cooldown_JHU} shows results from the first of a two-step process for the cooldown.  The first step was to understand the ambient heat load and thermal impedance  from the cooler to the thermal bars, irrespective of the optics.  The cryostat was pumped and cooled with no heater power applied to the thermal bars.  In this figure, the pressure is shown in the top panel and the temperature in the bottom panel.  

At roughly 3~x~$10^-7$~Torr, the pressure is well below the target of 1~microTorr.  This bodes well for the lifetime of the ion pumps.  The blip in the pressure plot at 14:30 is simply offgassing from ion pump startup; an artifact we have seen in other systems.  

\begin{figure}
    \centering
    \includegraphics[width=13.0cm]{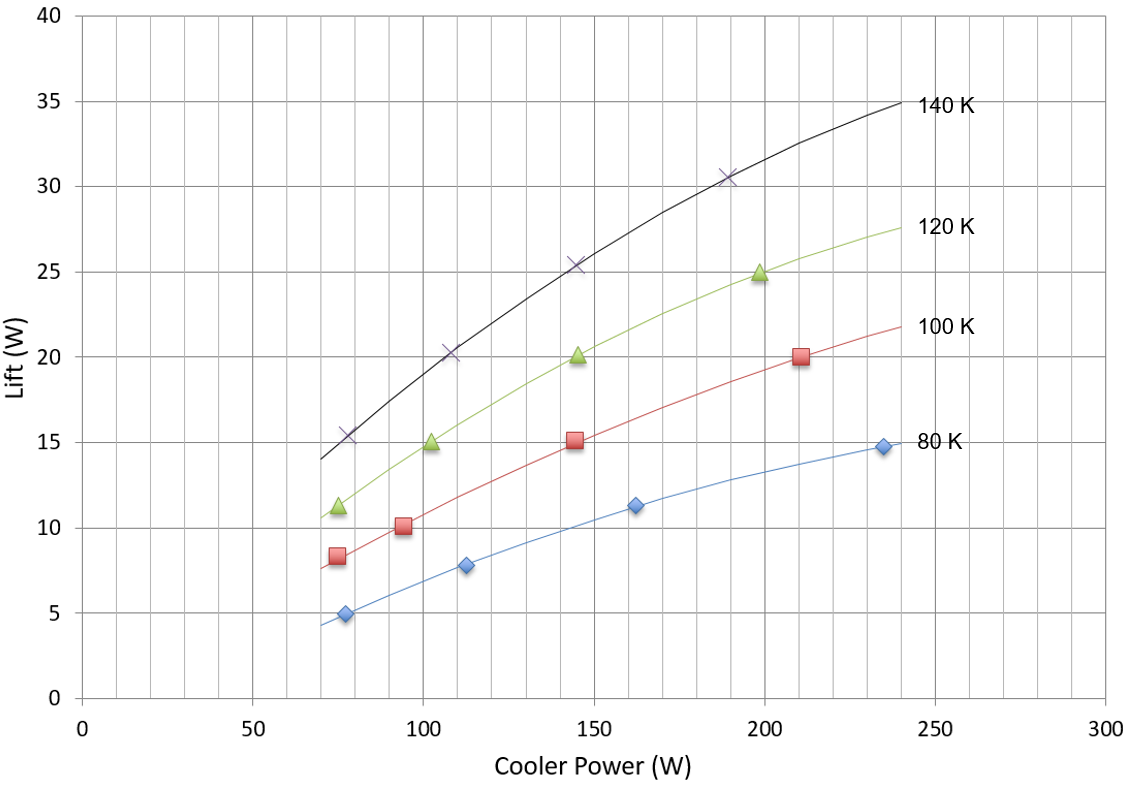}  
    \caption{Sunpower Cryotel GT measured performance (i.e. lift) as a function of tip temperature and cooler power.  Data points are measured values. Curves represent second-order polynomial fits to the data.}
    \label{fig:rossplot}
\end{figure}

The temperature/power plot shows that the cooler achieves the 100~K target with a power of 70~W, well below the maximum power of 250~W.  And the temperature of the thermal bar, 118~K, is well below the 140~K target, which one would expect with no thermal load applied. We glean from these early results that the total radiant and conductive thermal losses from the cryostat, without the optics, is about 7.5~W, which agrees very well with the estimate. This we know from measured performance of the cooler as a function of tip temperature and applied power; see Figure~\ref{fig:rossplot}.

\begin{figure}
    \centering
    \includegraphics[width=17.0cm]{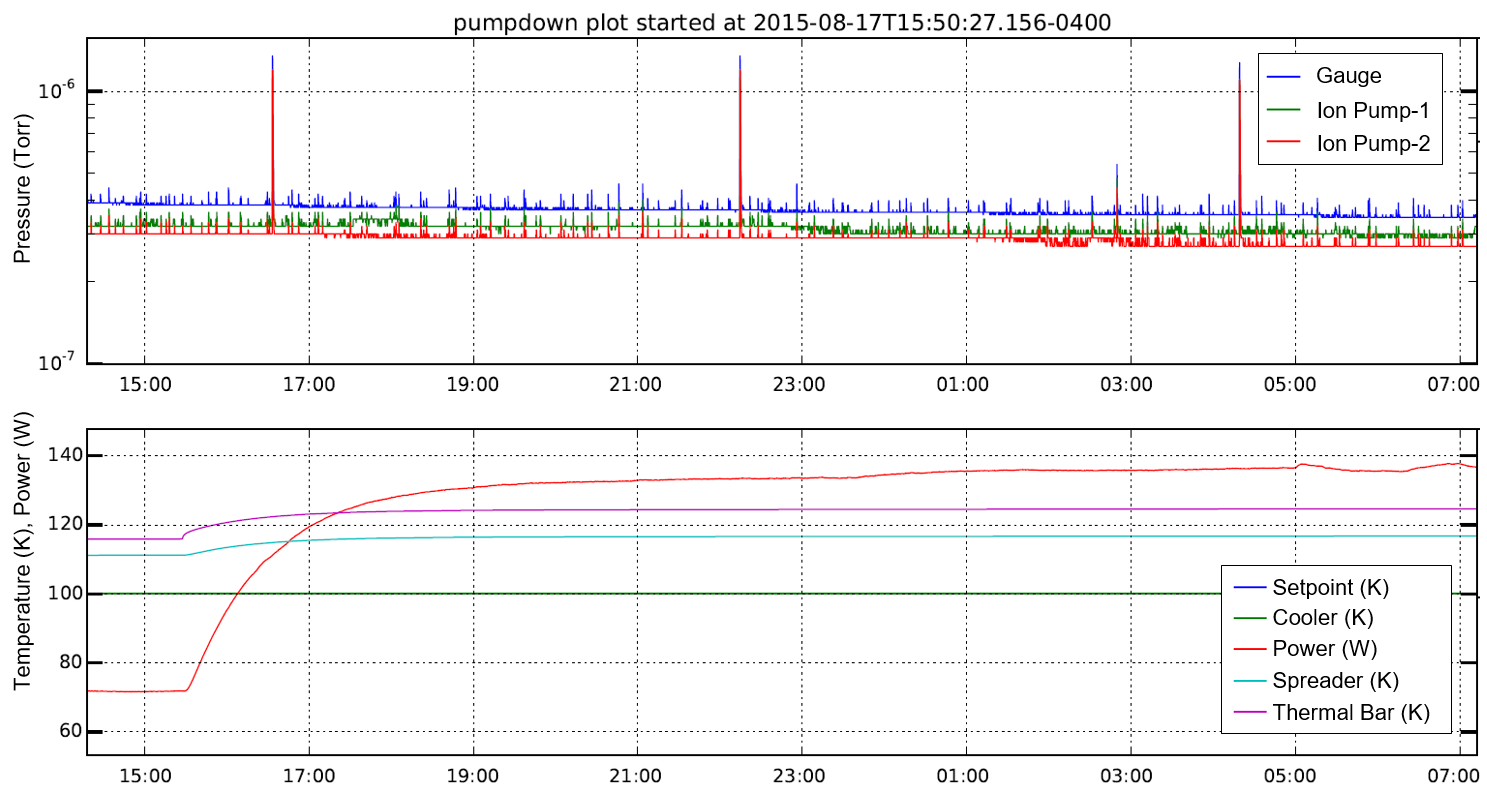}  
    \caption{Plot showing the cyrostat pressure (top panel), and temperatures and cooler power (bottom panel), for the first red cryostat with an applied heat load of 7~W to the ends of the thermal bars.  The added heat load simulates the ambient load induced by the presence of the optics.}
    \label{fig:heatload_JHU}
\end{figure}

In the second step of the cooldown, 7~W is applied to the thermal bars to simulate the ambient load produced by the presence of the optics.  Figure~\ref{fig:heatload_JHU} shows the results.  From the temperature/power plot, it can be seen that the cooler power increases to approximately 135~W, the temperature of the thermal spider increases to $\sim$118~K, and the thermal bar increases to $\sim$123~K, well below the 140~K target.  From the cooler performance curves we see that the total lift for 135~W cooler power is approximately 14~W, which agrees well with the total thermal load estimate of 14.5~W.

The post-shipment performance of the cryostat is discussed in Sec.\ref{sec:performance_LAM}.  However, there is one bit of discovery from that testing that is worth mentioning here.  During testing at LAM, it was discovered that the thermal calibration curve used in the temperature control module (see Sec.~\ref{sec:tempboard}) was incorrect. This affected the accuracy of the thermal spreader and thermal bar temperatures in Figures~\ref{fig:cooldown_JHU} and \ref{fig:heatload_JHU}; the cooler temperature, which is read by the cooler controller, was not affected.  The faulty calibration led to readings that were 4~K higher than actual, but only at low temperatures, $\sim$100~K. Hence, the steady state temperatures of the thermal spreader and thermal bar in Figure~\ref{fig:heatload_JHU} were actually 114~K and 119~K, respectively. And similarly, in Figure~\ref{fig:cooldown_JHU}, the steady state thermal spreader and thermal bar temperatures were 108~K and 114~K, respectively.

\subsection{Vacuum and Thermal Performance Tests at LAM}
\label{sec:performance_LAM}
\begin{figure}
    \centering
    \includegraphics[width=13.0cm]{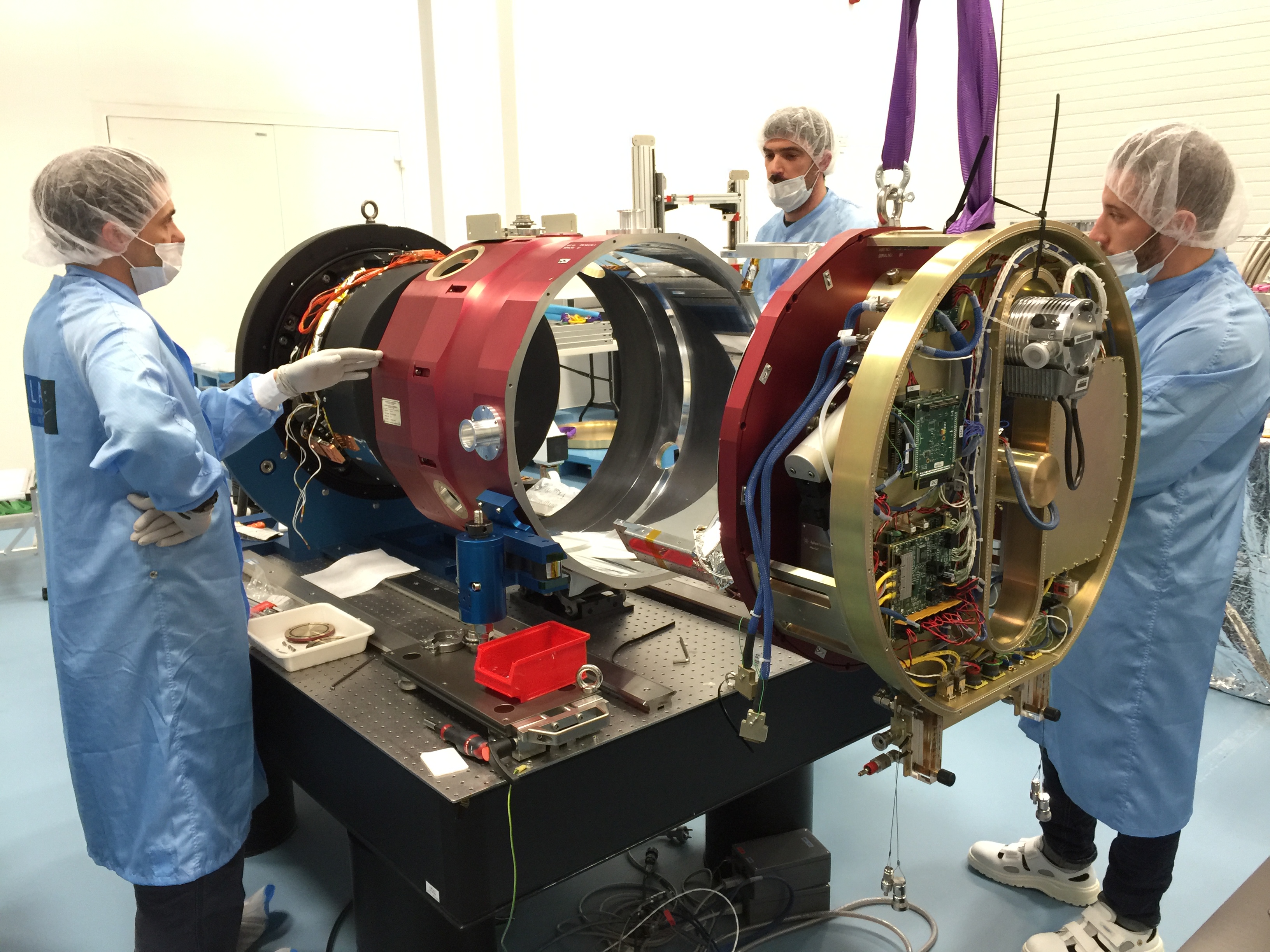}  
    \caption{Photograph of the red cryostat in the cleanroom at LAM.  In this photo, the main sections of the cyrostat have been separated in preparation for thermal testing.  The front bell and attached optics (in black) are visible in the background.  The rear cover, along with its associated thermal, vacuum, and electronics hardware are visible in the foreground.  Personnel in this composition give a sense of scale.}
    \label{fig:red_cryostat_LAM}
\end{figure}

After being received by the Laboratorie d'Astrophysique de Marseille and configured in their cleanroom (see Fig.~\ref{fig:red_cryostat_LAM}), the components from Winlight Systems were installed, along with additional thermal sensors for a system-level thermal validation test. Early results uncovered a problem.  The detector temperature could not be achieved, even with the cooler running at maximum power.  This set in motion a series of tests to debug the system.  At present tests are still being conducted.  So far, test results have uncovered some minor issues, which are discussed below, but the more serious issue, an apparent thermal short of order 10~W, has yet to be found.    

The first minor issue uncovered was that the temperature board was not calibrated properly.  This affected the temperature readings of the permanently installed thermal sensors (i.e. those not temporarily installed just for thermal validation), except the cryocooler tip sensor, which is read by the cryocooler controller.  This led to a 4~K error in temperature readings at $\sim$100~K, an error that gradually decreased toward room temperature; room temperature values were accurate. This problem was resolved.   

Another minor issue discovered was a 5~K temperature difference between the cooler tip and thermal collar, which resulted from a mismatch (of order 100$\mu$m) in diameter between the two mating surfaces.  A new cooler was installed when the performance of the original cooler was in question; the new cooler having a tip diameter that was better matched to the collar.  This solved the problem in the near-term.  The long term solution is to modify the collar such that it will conform to either a large or small diameter, a modification that is being worked now.

Once these two issues were resolved, we were able to reach the desired detector temperature of 163~K, but barely, and with the cryocooler running at full power, rather than an expected value of $\sim$135~W.

\begin{figure}
    \centering
    \includegraphics[width=13.0cm]{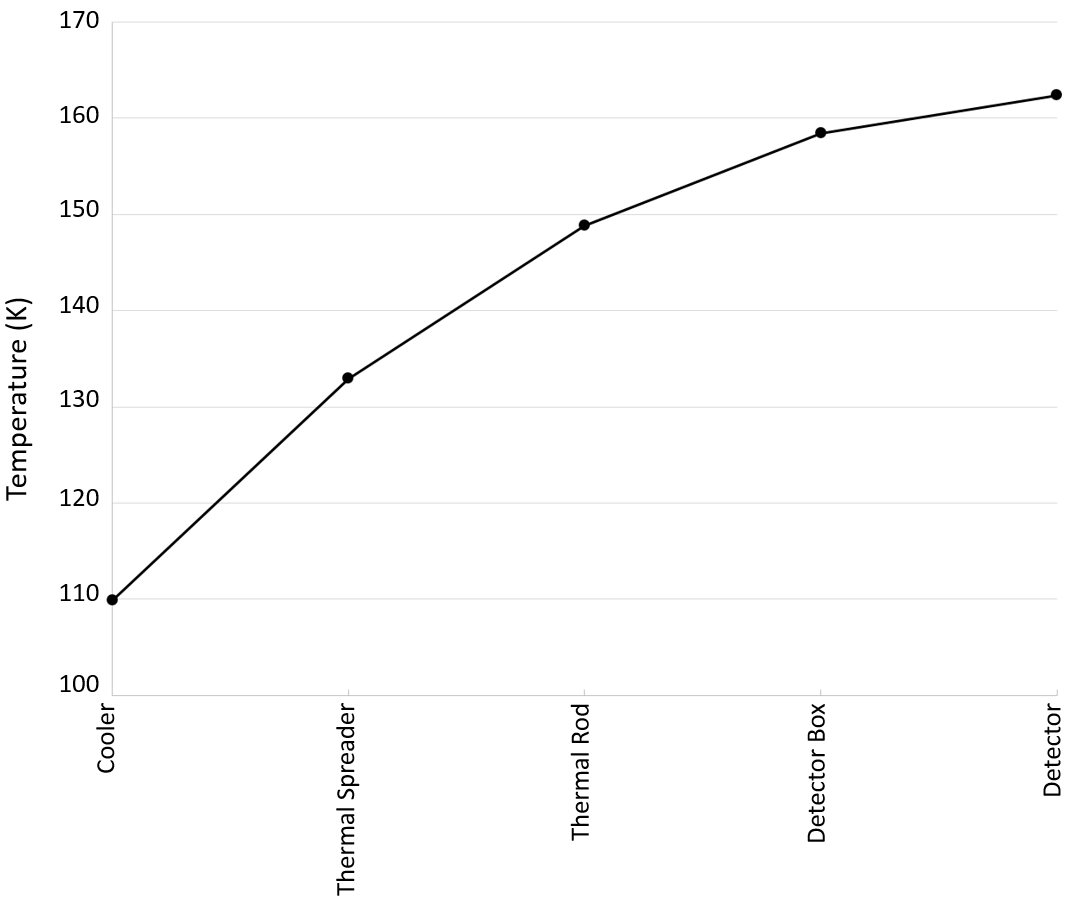}  
    \caption{Plot showing cryostat temperatures with the optics installed and the cooler at maximum power for a cooler setpoint of 100~K. At this cooler tip temperature the maximum cooler power is 230~W.  Note, given the heat load, the cooler is only able to reach 110~K.}
    \label{fig:cooldown_LAM}
\end{figure}

Recent tests results have all produced temperature profiles very similar to that shown in  Figure~\ref{fig:cooldown_LAM}.  This result is for a 230~W input power into the cryocooler, the maximum possible at that tip temperature. We should note that the maximum input power varies with temperature.    

Additional diagnostics have since ruled out a number of potential sources for the unexpected thermal loss.  Radiative load on the detector spider and detector box has been verified to agree with the initial estimate.  And a short between the thermal rod and ambient support hardware has pretty much been ruled out.  At the moment, the number of possible culprits is limited, and we are hopeful that we will solve this problem in the very near term. 

\subsection{Electrical Performance}
The electronics have performed well overall.  The system is robust and we have had no failures within the electronics mounted to the cryostat, or in the control rack.  

We did however, discover an issue with in-rush current burning out the MOSFETs on the cooler remote power switch, which led to a redesign of the board.  The problem was actually discovered during bench-testing, and not during cryostat qualification.  The bench test revealed the issue because that setup did not include the inductance of the long cable connecting the supply rack to the pie-pan.   

Another issue discovered, and corrected, was noise pickup by the TCM; the source being the power control module beneath the board.  This problem was solved by redesigning the TCM mounting frame to incorporate a metal plate to isolate the boards.

\section{Summary}
We have described the design of the PFS visible camera cryostats and have discussed their performance.  The as-built design has been shown to satisfy most requirements, and performs very well.  As is typical for complex engineering efforts such as this, there have been some issues, most of which have been minor and readily resolved.  As of this writing, the pressing issue facing us is the identification of a greater than expected thermal load on the cryostat, and of course resolving it.  But first we must find the problem.

\acknowledgments 
We gratefully acknowledge support from the Funding Program for World-Leading Innovative R\&D in Science and Technology (FIRST), program:
"Subaru Measurements of Images and Redshifts (SuMIRe)", CSTP, Japan

\bibliography{report} 
\bibliographystyle{spiebib} 

\end{document}